\documentclass[twocolumn]{aa}
\usepackage{graphicx}

\usepackage{txfonts}

\newcommand{\micron}{$\mu$m}

\begin{document}
   \title{The 2D Continuum Radiative Transfer Problem}

   \subtitle{Benchmark Results for Disk Configurations}

   \author{I. Pascucci\inst{1} 
                \and
           S. Wolf\inst{1,2}
                \and
           J. Steinacker\inst{1,3}
                \and            
           C. P. Dullemond\inst{4}
                \and
           Th. Henning\inst{1}
                \and            
           G. Niccolini\inst{5}
                \and
           P. Woitke\inst{6}
                \and
          B. Lopez\inst{5}}

   \offprints{I. Pascucci, pascucci@mpia-hd.mpg.de}

   \institute{Max-Planck-Institut f\"{u}r Astronomie,
              K\"{o}nigstuhl 17, D-69117 Heidelberg, Germany
         \and
              California Institute of Technology,
	      1201 East California Blvd, Mail Code 105-24, Pasadena, CA 91125, USA
         \and    
              Astrophysikalisches Institut und Universit\"{a}ts-Sternwarte,
              Schillerg\"{a}sschen 2-3, D-07745 Jena, Germany
         \and     
              Max-Planck-Institut f\"{u}r Astrophysik, 
              Karl-Schwarzschild-Str. 1, D-85741 Garching, Germany 
         \and
             Observatoire de la C\^{o}te d'Azur, D\'{e}partement Fresnel,
             UMR 6528, BP 4229, 06034 Nice Cedex 4, France
        \and
             Zentrum f\"ur Astronomie und Astrophysik, TU Berlin, 
             Hardenbergstra{\ss}e~36, D-10623 Berlin, Germany       
             }

   \date{Received xxx; accepted xxx}

   \abstract{
   We present benchmark problems and solutions for the continuum radiative
   transfer (RT) in a 2D disk configuration.
   The reliability of three Monte-Carlo and two grid-based codes is tested 
   by comparing their results for a set of well-defined cases which differ
   for optical depth and viewing angle.
   For all the configurations, the overall shape of the resulting temperature 
   and spectral energy distribution is well reproduced.
   The solutions we provide can be used for the verification of other RT codes.    
   We also point out the advantages and disadvantages of the various numerical 
   techniques applied to solve the RT problem.
   
   }

   \maketitle
%

\section{Introduction}

        Observations show that many astrophysical sources such as young 
        stellar objects (YSOs), post-AGB stars and active galactic nuclei 
        (AGN) are surrounded by dust.
        Dust grains scatter, absorb and re-emit radiation originating from 
        the primary energy sources, thus modifying their spectral energy 
        distributions (SEDs). 
        Moreover many embedded objects cannot be directly studied in the 
        visible, since dust may entirely obscure them at optical wavelengths.
        Their structure can be only inferred from the thermal dust emission.
        Therefore, modelling of their intensity and polarization maps as 
        well as their SEDs is necessary.
        This can only be done by solving the radiative transfer (RT) equation 
        (e.g. Yorke \cite{yorke85}). 
        Analytical solutions for this equation do exist only for the 
        simplest cases, far from representing the complexity of dust-enshrouded
        objects. Hence, the development of sophisticated numerical RT codes is 
        unavoidable.
        
        Early attempts for spherically symmetric configurations were performed 
        by Hummer \& Rybicki (\cite{hummer71}), Scoville \& Kwan 
        (\cite{scoville76}) and Leung (\cite{leung76}) including rough 
        assumptions such as grey opacity and/or neglecting scattering.
        The first formal solution for the dust continuum in spherical geometry 
	was obtained by Rowan-Robinson (\cite{rowan80}), directly integrating the RT equation, 
        an operation known as ray-tracing.
        Since then many other codes treating 1-D slab or spherical 
        configurations (e.g. Yorke \& Shustov \cite{yorke81}; 
        Lef\`evre et al. \cite{lefevre82}; Martin et al. \cite{martin84}; 
        Rogers \& Martin \cite{rogers84}; Henning \cite{henning85}; 
        Groenewegen \cite{groenewegen93}; Winters et al. \cite{winters94}) 
        or the inverse RT problem (Steinacker et al. \cite{stein02b}) have been developed.
        Nevertheless, it became soon clear that a 1-D geometry is often too 
        restrictive.
        Distinct non-spherical features such as bipolar outflows 
        (e.g. Bachiller \cite{bachiller96}), bipolar reflection nebulae 
        (e.g. Lenzen \cite{lenzen87}) and disks 
        (e.g. McCaughrean \& O'Dell \cite{mccau96}) are typical of many 
        astronomical objects.
         Nowadays,  multidimensional codes implementing 
        different methods and numerical schemes are being applied to treat 
        the RT in such configurations.
        
        In contrast to hydrodynamical simulations, benchmark tests for 
        radiative transfer computations are rare. 
        The only practical approach to test the reliability of RT calculations is to 
        compare solutions of well-defined problems by several independent codes.
        This has been done for the 1D case by Ivezic et al. (\cite{ivezic97}).
        A benchmark project for 1-D plane-parallel RT and vertical structure
        calculations for irradiated passive disks is available on the web
        \footnote{http://www.mpa-garching.mpg.de/PUBLICATIONS/DATA/radtrans/
        \hspace*{1.1cm}benchmarks/}.
        As for 1-D RT in molecular lines, a comparison of results 
        from different codes has been performed by van Zadelhoff et al. (\cite{vanzadelhoff02})
        \footnote{see also:
        http://www.strw.leidenuniv.nl/$\sim$radtrans/}.
        Going from spherical symmetry to 2D and 3D spatial configurations,
        we add two or three more variables to the RT problem.
        Numerically, this implies  10$^4$ or 10$^6$ more numbers to
        store when a decent resolution of 100 points in each variable
        is used. In addition, the geometry makes the solution of the
        integro-differential RT equation more complex. 
        This explains why benchmark tests for 2D and 3D configurations
        are lacking. It also implies that reaching an agreement to the
        level of 1D RT computations using state-of-the-art computer equipment 
        is unrealistic. 
        A previous attempt to test 2D RT calculations has been made
        by Men'shchikov \& Henning (\cite{mensh97}). They compare results from their 
        approximate method with those of a fully-2D program 
        (Efstathiou \& Rowan-Robinson \cite{efstathiou90}) applying the same 
        geometry.

        Here, we propose to test the behaviour of five different RT codes
        in a well defined 2D configuration, point out advantages and 
	disadvantages of the various techniques applied to solve the 
	RT problem and provide benchmark solutions for the verification 
	of continuum RT codes.
	As modelling sources with high optical depth and strong scattering is
	the challenge of multi-dimensional RT codes, we explicitly
	include a test case at the limit of the current computational
	capabilities.	
        In Sect.~\ref{sect:bprob} we briefly introduce the RT problem and
        we define our test case.
        Sect.~\ref{sect:codes} is devoted to the description of the codes 
        we used and to explain their differences.
        Solutions for the dust temperature and emerging SEDs are presented in
        Sect.~\ref{sect:bres}. In the last section we discuss our results. 

\section{Benchmark Problem \label{sect:bprob}}
\subsection{The RT problem}
        Solving the RT problem means to determine the
         intensity $I_\lambda(\vec x,\vec n)$
        of the radiation field at each point $\vec x$ and 
        direction $\vec n$ of the model geometry and at each 
        wavelength $\lambda$. 
        This is achieved by solving the stationary transfer equation
\begin{eqnarray}\label{radtrans}
\vec n \nabla_{\vec x}
           I_\lambda(\vec x,\vec n)
           & = &    
-\left[\kappa^{\mathrm{abs}}(\lambda,\vec x)\ + \kappa^{\mathrm{sca}}(\lambda,\vec x)\right]\
           I_\lambda(\vec x,\vec n) \nonumber \\
& &+ \kappa^{\mathrm{abs}}(\lambda,\vec x)\
           B_\lambda[T(\vec x)]   \nonumber \\     
& &+ \frac{\kappa^{\mathrm{sca}}(\lambda,\vec x)}{4\pi}
           \int\limits_{\Omega}d\Omega'
           p(\lambda,\vec n,\vec n')
           I_\lambda(\vec x,\vec n')  \nonumber \\        
& &+ E_\lambda(\vec x,\vec n)      
\end{eqnarray}
        where $\kappa^{\mathrm{abs}}(\lambda,\vec x)$ and $\kappa^{\mathrm{sca}}
        (\lambda,\vec x)$ are the absorption and scattering coefficients of the particles,
        respectively. The quantity $p(\lambda,\vec n,\vec n')$ denotes the 
        probability that radiation is scattered from the direction $\vec n'$ 
        into $\vec n$, $\Omega$ is the solid angle, $B_\lambda$ is the 
        Planck function, and $T$ is the temperature. 
        The index $\lambda$ denotes that the quantity is defined per 
        wavelength interval. 
        $E_\lambda(\vec x,\vec n)$ represents all internal radiation sources 
        such as viscous heating or cosmic rays. For the sake of simplicity,
        we only consider one dust component of specific size and chemical composition.
        In addition, we do not discuss the polarization state of the radiation field
        and consider the intensity only.
        
        If spherical symmetry in the particle distribution and the sources 
        of radiation is assumed, the integro--differential equation 
        (\ref{radtrans}) becomes a function of 3 variables, already difficult to solve 
        even for a given dust temperature $T(\vec x)$. In the case of spatial
        2D configurations (axial-symmetric disks, tori), we have to deal with 5
        variables.  Moreover, the coupling between the radiation field and the
        dust temperature requires the simultaneous consideration of the balance
        equation for the local energy density at point $\vec x$
\begin{equation}
\int\limits_0^\infty d\lambda\ Q^{\mathrm{abs}}_\lambda
   B_\lambda[T_{\mathrm{rad}}(\vec x)]  =
\int\limits_0^\infty d\lambda\ Q^{\mathrm{abs}}_\lambda
   \frac{1}{4\pi}
   \int\limits_\Omega d\Omega'\
   I_\lambda(\vec x,\vec n') \label{enerbal}
\end{equation}
        to calculate intensity and temperature self--consistently. Here,
        $Q^{\mathrm{abs}}(\lambda)$ is the absorption efficiency factor, 
        while $T_{\mathrm{rad}}$ is the temperature arising from radiative heating. 
        Additional heating sources can contribute to the temperature with
\begin{equation}
  T(\vec x)=T_{\mathrm{rad}}(\vec x)+T_{\mathrm{heat}}(\vec x).
\end{equation}
%
        
\subsection{Model definition}\label{defmodel}
\begin{figure}
        \resizebox{\hsize}{!}{\includegraphics[angle=0]{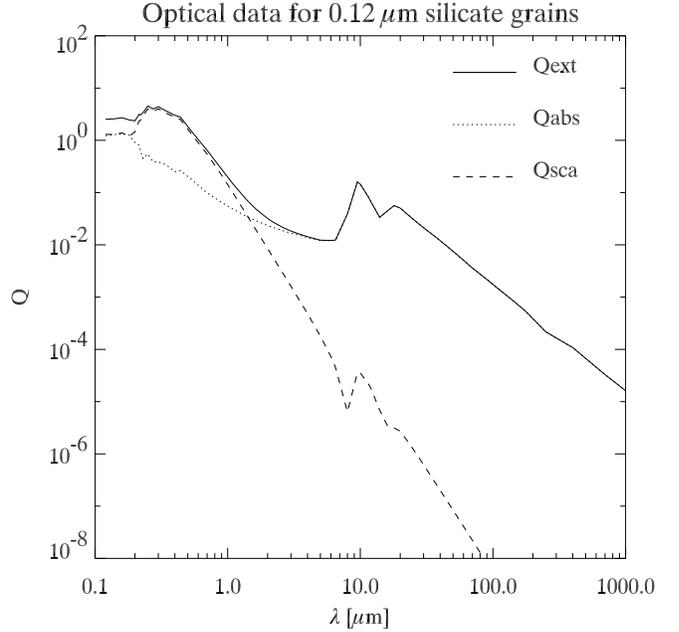}}
        \caption[]{
        Optical data for spherical astronomical silicate grains 
	having a radius of 0.12~\micron{} (Draine \& Lee \cite{draine84}).
        Note that scattering dominates between 0.2 and 1 \micron{} 
	for this type of grains. 
        } 
        \label{fig:opt}
\end{figure}

\begin{figure}
        \resizebox{\hsize}{!}{\includegraphics[angle=90]{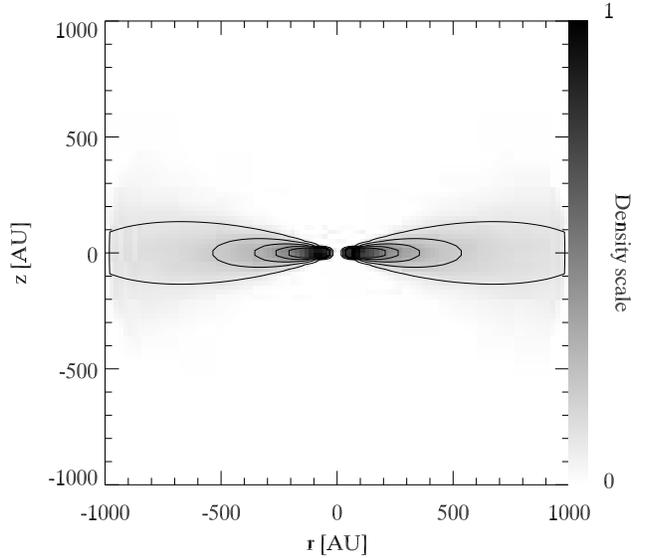}}
        \caption[]{
        Density structure perpendicular to the disk midplane and centered on the 
        star for the model with $\tau_{\rm v}$ = 1. 
        Values are normalized to the maximum density. 
        The contours provide 0.10, 0.19, 0.28, 0.38, 0.48\% of the maximum.
        } 
        \label{fig:den}
\end{figure}

        We consider the general astrophysical case of a star embedded in a 
        circumstellar disk with an inner cavity free of dust.
        We assume that the star is point-like, located at the center of the 
        configuration and radiating as a black body at the same temperature 
        as the Sun.
        The disk is made of spherical astronomical silicate grains, having a radius 
        of 0.12~\micron{} and a density of 3.6~g$\,$cm$^{-3}$
        (optical data are taken from Draine \& Lee 
        \cite{draine84}\footnote{downloadable from:\\ 
        http://www.mpia.de/PSF/PSFpages/RT/benchmark.html}, see also Fig.~\ref{fig:opt}).
        The disk radially extends to a maximum distance of 1000~AU from 
        the central star.  
        Since the correct determination of the sublimation radius is quite a 
        difficult problem, we fix the inner radius to 1~AU.
        This guarantees a maximum dust temperature less than 1000~K, even in 
        the case of high optical depth. 
        The density structure is that of a massless (in relation to the 
        central star) Keplerian disk having no cutoff at a certain opening 
        angle.
        This implies that the radiative transfer has to be simulated both         
	in the optically thick disk and in the optically thin envelope.        
        The disk geometry and density structure are similar to those described
	by Chiang \& Goldreich (\cite{cg97},\cite{cg99}) and successfully applied to 
	study passive disks around T Tauri stars (Natta et al. \cite{natta00}).
        The density distribution provides a steep-density gradient in the inner 
        part of the disk which could give rise to numerical problems when 
        solving the RT equation. 
        This turns out to be an advantage for RT comparison since it allows to 
        test the codes' behaviour under extreme conditions.
        The density distribution we adopt has the following form 
\begin{eqnarray}\label{eq:den}
\rho(r,z) & = & \rho_{\mathrm{0}} \times f_{\mathrm{1}}(r) \times f_{\mathrm{2}}(z/h(r)) \\
f_{\mathrm{1}}(r) & = & ( r / r_{\mathrm{d}} )^{-1.0}  \nonumber \\
f_{\mathrm{2}}(r) & = & exp( -\pi/4 \times (z/h(r))^2)  \nonumber \\
h(r)   & = & z_{\mathrm{d}} \times  (r / r_{\mathrm{d}})^{1.125}  \nonumber
\end{eqnarray}
        with $r$ being the distance from the central star in the disk midplane 
        ($\sqrt{x^2+y^2}$) and $z$ the distance from the midplane. 
        Here $r_{\mathrm{d}}$ is half of the disk outer radius (R$_{\mathrm{out}}$/2) and 
        $z_{\mathrm{d}}$ one fourth of  $r_{\mathrm{d}}$ (R$_{\mathrm{out}}$/8). 
        Note that the disk is slightly flared, i.e., the disk opening angle 
        $h(r)/r$ is exponentially increasing with the distance from the star.
        The term $f_1$ provides the radial dependence of the density 
        distribution. In protoplanetary disks, the volume density is usually
        proportional to $r^{-\alpha}$ with $\alpha$ in the range
        (1.8$\div$2.8) 
        (e.g. Wood et al. \cite{wood02} and Cotera et al. \cite{cotera01}).
        For this benchmark we use $\alpha$ = 1 in order to save CPU time.
        Both $f_{\mathrm{1}}$ and $f_{\mathrm{2}}$ remain unchanged while $\rho_0$ is 
        chosen so to define different optical depths. 
        We perform calculations for four values of visual ($\lambda$ = 550 nm) 
        optical depth, namely $\tau_{\mathrm{v}}$~= 0.1, 1, 10, 100. 
        The optical depth, as seen from the centre, is calculated along the 
        disk midplane. 
        Since most of the dust is confined in the midplane, 
        the optical depths we refer to are the highest in each model.
	The test case $\tau_{\mathrm{v}}$~=~100 is at the limit of our current
	computational capabilities.
        The resulting total dust mass for the model with $\tau_{\mathrm{v}}$~=~1(100)   
        is 1.1$\times$10$^{-6}$~M$_\odot$(1.1$\times$10$^{-4}$~M$_\odot$). 
        The density structure perpendicular to the disk midplane is shown for the 
        same model in Fig.~\ref{fig:den}.
        The RT  is calculated for 61 wavelengths being 
        distributed nearly equidistantly on a logarithmic scale from 
        0.12--2000~\micron . These 61 wavelengths
        define the frequency resolution of our computations.
        In Sect.~\ref{sect:tests} we also compare two Monte Carlo
	(MC) codes on a grid with
	two times more wavelengths and we discuss the effect of 
	the frequency resolution on the 2D benchmark.
        Since anisotropic scattering is not included in all codes, we 
        consider the scattering as isotropic.
        Symbols and values of the model parameters are summarized in
        Table~\ref{tab:param} for more clarity. 
        
   \begin{table}
      \caption[]{Model parameters}
         \label{tab:param}
     $$ 
         \begin{array}{p{0.15\linewidth}p{0.5\linewidth}p{0.25\linewidth}l}
            \hline
            \hline
            \noalign{\smallskip}
            Symbol              & meaning                       & value \\              
            \noalign{\smallskip}
            \hline
            \noalign{\smallskip}
            $M_*$               & Stellar mass                  & 1~M$_\odot$           \\
            $R_*$               & Stellar radius                & 1~R$_\odot$           \\
            $T_*$               & Stellar effective temperature & 5800~K                \\
            $R_{\mathrm{out}}$  & Outer disk radius             & 1000~AU               \\
            $R_{\mathrm{in}}$   & Inner disk radius             & 1~AU                  \\  
            $z_{\mathrm{d}}$    & Disk height                   & 125~AU                \\
            $a$                 & Grain radius                  & 0.12~\micron          \\ 
            $\rho_{\mathrm{g}}$ & Grain density                 & 3.6~g$\,$cm$^{-3}$     \\ 
            $\tau_{\mathrm{v}}$ & Optical depth at 550~nm       & 0.1, 1, 10, 100        \\               
            \noalign{\smallskip}
            \hline
         \end{array}
     $$ 
   \end{table}  

\section{RT Simulations}{\label{sect:codes}}
\subsection{Methods}
        Similar to hydrodynamical simulations, we can distinguish particle
        (Monte Carlo) and grid-based methods to solve the RT equation numerically
        (Henning \cite{henning01}).

        In MC simulations the radiation field is partitioned in equal-energy,
        monochromatic "photon packets" that are emitted stochastically both
        by the source and by the surrounding envelope.  The optical depth
        determines the location at which the packets interact while their
        albedo defines the probability of either scattering or absorption.  In the
        original scheme ({\it scheme 1}) the source and the envelope photon
        packets are emitted separately.  At first the grains re-emit according
        to the absorbed source radiation.  Then dust reemission takes
        also into account the envelope emission radiation field. Reemission by
        the dust is repeated as long as the difference between the input and
        the output energy is larger than a chosen threshold. However, the dust
        reemission, i.e. the repetition of the Monte Carlo experiment, is time
        consuming. An alternative possibility ({\it scheme 2}) is to store
        all radiation exchanges within the envelope. In this case the Monte
        Carlo experiment can be carried out once for all\footnote{This is only
        valid for opacities explicitely independent on the temperature}, 
        but a large amount of computer memory is needed.
        A drawback of these two schemes is that the input luminosity is
        not automatically conserved during the simulation. This becomes a
        serious problem for configurations with very high optical depths which
        therefore usually need a larger number of iterations.  A solution has
        been found by Bjorkman \& Wood (\cite{bjork01}) in the so-called
        immediate reemission ({\it scheme 3}).  In this case only source
        photon packets are emitted and followed in their interaction
        locations. When a packet is absorbed, its energy is added to the
        envelope and a new packet is emitted immediately at a frequency which
        takes into account the modified envelope temperature.  This method
        does not require any iteration and implicitly conserves the total
        energy.  Another improvement  of the standard MC procedure has been 
        proposed by Lucy (\cite{lucy99}) to treat extremely
        optically thin configurations.  This approach considers the absorption
        not only at the end points of the photon path but also in between.
   \begin{table*}
      \caption[]{Main features of the codes.}\label{tab:feature}
      $$ 
         \begin{array}{p{0.3\linewidth}p{0.12\linewidth}p{0.12\linewidth}p{0.12\linewidth}p{0.12\linewidth}p{0.12\linewidth}}
            \hline
            \hline
            \noalign{\smallskip}
            Feature                               & MC3D   &MCTRANSF&RADICAL     &RADMC           &STEINRAY     \\           
            \noalign{\smallskip}
                                                  &        &        &            &                &               \\
             \hline
            \noalign{\smallskip}
            3D                                    & +      &        &            &                & +           \\
            anisotropic scattering                & +      & +      &            & +              & +         \\
            arbitrary grid geometry               & +      & +      &            &                &            \\
            grain size distribution               & +      & +      & +          & +              & +         \\
            multiple dust species                 & +      &        & +          & +              & +         \\
            images                                & +      & +      & +          & +              & +         \\
            polarization maps                     & +      &        &            &                &            \\
            global error control                  &        &        & +          &                & +         \\
            multiple/extended heating sources     & +      &        & +          &                & +         \\
            dust evaporation included             &        & +      & +          &                &            \\
            acceleration for high $\tau$ (ALI/Ng/other) & +& +      & +          &                & +         \\
            parallel version                     &         & +      & +          &                &            \\      
            \noalign{\smallskip}      
            \hline
           \end{array}
     $$    
   \end{table*} 
   
        Grid-based codes solve the RT equation on a discrete spatial grid.
        The grid can be either determined during the simulation
        or generated before starting the computation. The 2 RT grid-based
        codes we compare, namely RADICAL and STEINRAY, use the second
        approach (see Sect.~\ref{RADICAL} and \ref{STEINRAY}).  Among the
        schemes applied to solve the RT problem two are the most used: the
        so-called "Lambda Iteration" (see e.g.~Collison \& Fix
        \cite{collfix91}; Efstathiou \& Rowan-Robinson \cite{efstathiou91})
        and the "Variable Eddington Tensor" (Mihalas \& Mihalas
        \cite{mihalmihal84}; Malbet \& Bertout \cite{malbetbertout91}; Stone et al.
        \cite{stonemihnor92}; Kikuchi et al. \cite{kikuchi02}; 
        Dullemond et al. \cite{dulvzadnat02}; Dullemond \cite{dullemond03}).  
        The "Lambda Iteration" mode suffers from the same convergence problems as the
        standard MC which is based on {\it scheme 1}.  Thus, the improved
        "Accelerated Lambda Iteration" (e.g.~Rybicki \& Hummer \cite{ryb91})
        method is more widely applied.  The "Variable Eddington Tensor" mode
        is more robust than the "Lambda Iteration" and usually converges
        faster. Moreover, it has been proven that it works properly even at
        extremely high optical depths.  Integration of the formal RT equation can be
        done in a straightforward way by applying the "Long Characteristics"
        algorithm.  This method is accurate but turns out to be costly in CPU
        time.  A more efficient way is based on the "Short Characteristics"
        algorithm (Mihalas et al. \cite{mihalas78}; Kunasz \& Auer
        \cite{kunasz88}).
 
        Each of the solution algorithms has its advantages and
        drawbacks. In MC methods, a photon is propagated through the
        calculation domain and its scattering, absorption, and re-emission are
        tracked in detail. This allows to treat very complicated spatial
        distributions, arbitrary scattering functions and polarization.
%
	Drawback is the presence of a random noise in the results. 
	This noise can be reduced by increasing the number of used 
	photon packages and by including deterministic elements
	in the MC experiment (Niccolini et al.~\cite{niccolini}). 
        Grid-based solvers are less flexible than MC codes but have 
	the advantage not to involve random noise.

\subsection{Code description}
In the following sections we briefly describe the RT codes  
participating in the 2D benchmark.
A summary of their main features is provided in Table \ref{tab:feature}.
\subsubsection{MC3D}\label{MC3D}
        MC3D is a 3D continuum RT code.  It is based on the MC method and
        solves the RT problem self-consistently.  MC3D is designed for the
        simulation of dust temperatures in arbitrary dust/electron
        configurations and the resulting observables: spectral energy
        distributions, wavelength-dependent images and polarization maps.

        For the estimation of temperatures either the standard {\it scheme 1}
        (see Wolf et al. \cite{wolf99b}; Wolf \& Henning \cite{wolf00}) or the
        immediate reemission concept ({\it scheme 3}) can be applied.  For
        this benchmark project, the {\it scheme 3} is used to treat properly 
        the more optically thick models.  Optically very thin configurations,
        such as the atmosphere/envelope described in Sect.~\ref{defmodel}, are
        easily computed by the method proposed by Lucy (\cite{lucy99}).  Furthermore,
        the efficiency of MC3D is increased by (a) the fast photon transfer
        and (b) wavelength range selection concept (see Wolf \& Henning
        \cite{wolf00} for details), and (c) the enforced scattering mechanism
        as described by Cashwell \& Everett (\cite{cashwell59}).

        Previous applications of MC3D cover feasibility studies of extrasolar
        planet detections (Wolf et al. \cite{wolf02}), the RT in the clumpy
        circumstellar environment of YSOs (Wolf et al. \cite{wolf98}), 
        polarization studies of T~Tauri stars (Wolf et al. \cite{wolf01a}a), 
        AGN polarization models (Wolf \& Henning \cite{wolf99a}), 
        a solution for the multiple scattering of polarized
        radiation by non-spherical grains (Wolf et al. \cite{wolf02}), and the
        inverse RT based on the MC method (Wolf \cite{wolf01b}b).
        Executables of MC3D\,(V2) can be downloaded for several model
        geometries and platforms from: \\
        http://www.mpia-hd.mpg.de/FRINGE/SOFTWARE/mc3d/ (current US mirror
        page: http://spider.ipac.caltech.edu/staff/\\swolf/mc3d/).
\subsubsection{MCTRANSF}\label{MCTRANSF}
        MCTRANSF solves multi-dimensional continuum RT problems in dusty media
        by means of a MC method. It has been originally developed by Lopez et
        al.~(\cite{lopez95}). So far, the code has been used for the empirical
        modelling of several circumstellar envelopes of post AGB-stars of
        different types (e.g.  Lopez \& Perrin~\cite{lopez00}), including
        multi-scattering effects.

        Currently, only spherical symmetric (1D) and axisymmetric (2D)
        problems can be considered, but an extension to 3D is possible and
        straightforward.  Several improvements of the standard MC procedure
        have been recently included (Niccolini et al. \cite{niccolini}) in order 
        to avoid the usual increase of the noise level which typically occurs in 
        extremely optically thin or
        optically thick situations. The concept suggested by Lucy (\cite{lucy99})
        is implemented to treat very optically thin cases.  
        Optically thick configurations are tackled
        by the inclusion of several deterministic elements for the treatment
        of the absorption during the photon propagation phase, forcing the
        absorption to take place all along the rays.  The temperature
        structure of the medium in radiative equilibrium is found by applying 
        {\it scheme 2}.  The convergence is found to be rapid, even in
        optically thick situations, but needs a large amount of computer
        memory, because the primary MC information must be stored
        source-dependently.

        As a result of the combination of all these measures, MCTRANSF is
        capable to simultaneously model optically thin and optically thick
        parts of the model volume with about the same accuracy. Thus, both
        clumpy media and discontinuous opacity structures can be handled.
        MCTRANSF is able to arrive at numerical solutions for RT problems even
        in case of very large optical depths (e.g. for disk configurations).
        Parallelised versions of the code have been developed for a Cray T3E
        1200 and for systems supporting the OpenMP application program
        interface. All these versions use shared memory systems.
\subsubsection{RADICAL}\label{RADICAL}
        The core of the code {\tt RADICAL} is a lambda operator subroutine
        based on the method of ``Short Characteristics'', implemented on a
        polar grid by Dullemond \& Turolla (\cite{dullemond00}). Using this
        subroutine as the main driver, {\tt RADICAL} offers two modes of
        operation: a simple Lambda Iteration mode and a Variable Eddington
        Tensor mode. In this paper we use the latter because of its faster
        convergence and capability of treating high optical depths.  The
        Variable Eddington Tensor method is implemented in RADICAL as follows
        (Dullemond \cite{dullemond03}).  First, the primary stellar radiation
        field is propagated from the star outwards into the disk. Dust
        scattering is included in a MC fashion.  The energy absorbed by the
        disk in each grid-cell is then re-emitted as infrared (IR) radiation, which
        is treated as a separate radiation field. The 2-D transfer solution
        for this secondary radiation field is found by solving the
        frequency-integrated moment equations. The closure for these equations
        is based on the variable Eddington tensors and mean opacities computed
        with the Short Characteristics method of Dullemond \& Turolla. At the
        end of the calculation a global check on flux conservation is made.
        For all the models discussed in this paper the error remained within
        2\%.
\begin{figure}
        \resizebox{\hsize}{!}{\includegraphics[angle=90]{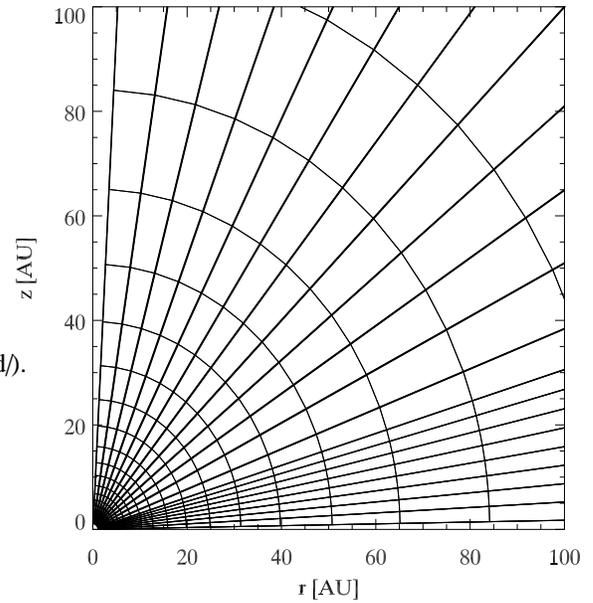}}
        \caption[]{Grid adopted by MCTRANSF to store the temperature resulting from
	the RT simulations. Similar spherical grids are also used by the other codes.
	} 
        \label{fig:polargrid}
\end{figure}
\subsubsection{RADMC}\label{RADMC}
        RADMC is an MC code based on {\it scheme 3}.  However, the original
        method of Bjorkman \& Wood produces very noisy temperature profiles in
        regions of low optical depth, and requires a large number of photons
        ($N\sim 10^7$) for a smooth SED. These disadvantages have been solved
        in RADMC by treating absorption partly as a continuous process
	(Lucy \cite{lucy99} ), and
        using the resulting smooth temperature profiles with a ray-tracing
        code to produce images and SEDs. These images and SEDs have a low
        noise level even for relatively few photon packages ($N\sim 10^5$). In
        addition, the frequency grid used for RADMC is not bound by the
        constraints set in the original method. 
        This improved Bjorkman \& Wood method works well at
        all optical depths, but may become slow in cases where the optical depth
        is very large ($\tau_{\mathrm{v}}$ about 1000). 
	For the test cases in this paper the optical depths are 
	low enough that this problem does not
        play a role.  
	For more information on the code, see {\tt
        http://www.mpa-garching.mpg.de/PUBLICATIONS/DATA/} {\tt
        radtrans/radmc/}.
   \begin{table*}
      \caption[]{Resolution and number of photons for the different test cases.}
         \label{tab:simu}
     $$ 
  \begin{array}{p{0.2\linewidth}p{0.05\linewidth}p{0.15\linewidth}p{0.05\linewidth}p{0.12\linewidth}p{0.11\linewidth}p{0.13\linewidth}l}
            \hline
            \hline
            \noalign{\smallskip}
          Code    & \# r&$\Delta$ r  & \# $\theta$& $\Delta$ $\theta$  &\# Phot & Test case        \\
          Name    &     &  [AU]      &            &[ $\degr$ ]      & [$\times$10$^6$] & $\tau_{\mathrm{550\rm{nm}}}$\\
            \hline
            \noalign{\smallskip}
          MC3D     & 55 & 0.03--141 &  121      & 1.5          & 244  & 0.1,1,10 \\
          MC3D     &10$^3$& 0.07--4.1 &  121    & 1.5          & 244   & 100 \\
          \hline   
          MCTRANSF & 48 & 0.17--125 &  40       & 4.5          &   1000  & 0.1 \\
	  MCTRANSF & 48 & 0.17--125 &  40       & 4.5          &   800   &  1 \\
          MCTRANSF & 46 & 0.18--130 &  46       & 2.8-5.3       &  1000  & 10 \\
	  MCTRANSF & 46 & 0.18--130 &  46       & 2.8-5.3       &  500   & 100  \\
          \hline 
          RADICAL  & 60 & 0.03--116 & 62        & 1.6-8.3       &       &0.1--100\\
          \hline 
           RADMC   & 60 & 0.03--116 & 62        & 1.6-8.3       & 10   &0.1--100          \\
          \hline 
           STEINRAY& 61 & 0.12--109 & 61        & 1.3           &       &0.1--100   \\ 
            \hline
         \end{array}
     $$ 
   \end{table*}      	
\subsubsection{STEINRAY}\label{STEINRAY}
        STEINRAY is a grid-based code which solves the full 3D continuum RT
        problem. A combination of ray-tracing and finite differencing of 2nd
        order on adaptive multi-frequency photon transport grids is applied.
        Steinacker et al. (\cite{stein02a}) have shown that the use of 3rd
        order finite differencing is too time-consuming for 3D RT, while 1st
        order schemes introduce an unacceptable degree of numerical diffusion
        to the solution.

        The spatial grids are generated using an algorithm described in
        Steinacker et al. (\cite{stein02c}).  They are adaptive and optimized
        to minimize the 1st order discretization error hence guaranteeing
        global error control for solutions of radiative transfer problems on
        the grid.  Since the use of one single grid for all frequencies leads
        to large discretization errors, STEINRAY calculates individual grids
        for each frequency to use the global error control of the grid
        generation method.  Minimization of the grid point number is possible
        in regions where the optical depth becomes large allowing for
        treatment of applications with optical depth of any value.
        Contrary to former treatments, the full frequency-dependent problem is
        solved without any flux approximation and for arbitrary scattering
        properties of the dust.
        For the direction discretization, equally spaced nodes on the unit
        sphere are used along with corresponding weights for the integration
        derived by evaluating special Gegenbauer polynomials in Steinacker et
        al. (\cite{stein96}).
        The temperature distribution is calculated by an Accelerated Lambda
        Iteration between the radiative transfer equation and the local balance
        equation. 
        The program is designed to provide spatially resolved images and 
        spectra of complex 3D dust distributions and allows for multiple 
        internal and external sources (Steinacker et al. \cite{stein03}).

        Recently, a 2D version of the program has been developed and was used for
        this benchmark. The grids are similar to the spherical grids shown in
        Fig. \ref{fig:polargrid}. The 2D version uses ray-tracing to solve for the intensity in all
        directions.
\subsection{Reliability of the codes in 1D}
        All the RT codes participating in the benchmark have been already
        tested in 1D spherically symmetric configurations.       
        Results from MC3D have been compared with those calculated
        with the RT code of Chini et al. (\cite{chini86})
        and with the code of Menshchikov \& Henning (\cite{mensh97}).
        In both cases differences below 1\% 
        even in the case of high optical depth have been found 
        (Wolf et al. \cite{wolf99b}).
        MCTRANSF has been tested (Niccolini et al. \cite{niccolini}) 
        against the 1D code written by Thibaut le Bertre and based on 
        the work of Leung (\cite{leung76}). 
        For the case of optical depth equal to 10 at
        1 \micron{} (with $\kappa \propto \frac{1}{\lambda}$, and isotropic
        scattering) the agreement between MCTRANSF and le Bertre's code is better than 1\%.
        In the case of  RADICAL, tests have been performed by comparing the results
        with those from 
        TRANSPHERE\footnote{http://www.mpa-garching.mpg.de/PUBLICATIONS/DATA/
        \hspace*{1.1cm}radtrans/},              
        a 1-D variable eddington factor code tested against the similar "code 1" 
        of Ivezic et al. (\cite{ivezic97}).
        Steinacker et al. (\cite{stein03}) used the 1D benchmark provided by 
        Ivezic et al. (\cite{ivezic97}) to test the 3D version of STEINRAY. 
        Agreement below 1\% has been found both for the emerging temperature and SEDs.        
\subsection{Details on the 2D RT computation}\label{sect:details} 
    The three MC codes involved in the 2D benchmark, namely MC3D (Sect.~\ref{MC3D}) and
    MCTRANSF (Sect.~\ref{MCTRANSF}) and RADMC (Sect.~\ref{RADMC}),
    choose a spherical grid to store the temperature resulting from the RT
    simulations. Radially the steps are logarithmic in order to properly
    resolve the innermost dense region of the disk. The number of radial
    points is kept below 100 for all the codes but MC3D, for which we tried
    a two weeks long computation for the most optically thick case 
    (see Table~\ref{tab:simu}). Differences in the SED between this computation and
    the one with 55 radial points and lower number of photons 
    are discussed in Sect.~\ref{sect:tests}. 
    MC3D adopts an equally spaced grid in vertical
    direction (1.5$\degr$ resolution, see Table~\ref{tab:simu}), while
    MCTRANSF and RADMC choose a resolution decreasing with the distance from the disk midplane
    (see Fig.~\ref{fig:polargrid}).
    Similar grid geometries are also used by the two grid-based codes RADICAL
    and STEINRAY.
    RADICAL (Sect.~\ref{RADICAL}) makes use of the same grid as RADMC
    while STEINRAY (Sect.~\ref{STEINRAY}) has comparable resolution in
    vertical direction but larger cells in the inner disk (0.12~AU, see
    Table~\ref{tab:simu}).
    
   We note that the resolution given in Table~\ref{tab:simu}, together with the number of photons
   for the MC codes, are at the limit of the computing capabilities for most of the codes.   
   MC3D needs about 1~Gby memory. The temperature resulting from these test cases is obtained in 1--2 days.
   Computing the SED requires about a week for all the models, but for the most optically thick one for which
   we try a longer computation.
   MCTRANSF needs a large amount of computer memory to store all radiation exchanges.
   The necessary memory goes as the square of the number of
   cells: $46\times 46$ cells is actually the technical limit ($\sim 4 \rm Gby$) 
   on our 4-processors (ev67 at $1\,\rm GHz$) HP-compaq ES45
   server. The runtime for the most optically thick case is about 2 weeks.
   Results from RADICAL and RADMC are obtained in less than a day for all the test cases.
   However, the actual spatial resolution of RADICAL cannot be doubled due 
   to not sufficient computer memory. 
   To produce the final spectrum RADMC uses the ray-tracing module from 
   RADICAL, and for that reason is also limited to the maximum spatial
   resolution that can be achieved by RADICAL. This is a mere technical
   problem.
   In the 2D version of STEINRAY, the code uses about 1~Gby memory. To
   obtain convergence in the temperature iteration of 1\% for the case
   $\tau_{\mathrm{v}}$~=~100, the code runtime is about a week. 
\section{Results \label{sect:bres}}

   \begin{figure}
   \resizebox{\hsize}{!}{\includegraphics[angle=90,width=2.5cm,height=2.5cm]{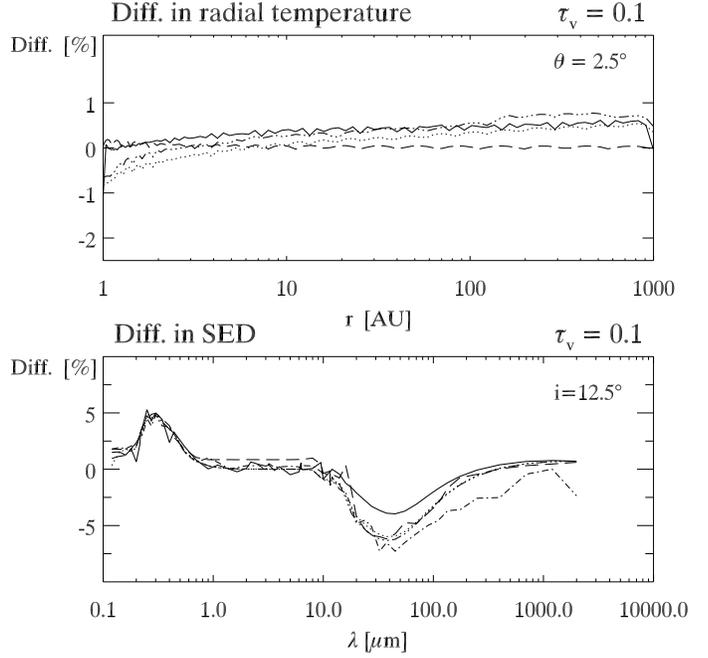}}
      \caption{
      Differences of the most optically thin model from the semi-analytical solution 
      (see Sect. \ref{sect:semi-ana}). Upper panel: differences in radial temperature for an angle $\theta$
      near to the disk midplane. Lower panel: differences in the SED for an almost face-on disk 
      (disk inclination equal to 12.5$\degr$). 
      For both panels, solid lines give the difference of MC3D, dot-dashed lines of MCTRANSF, 
      dashed-dot-dot-dot of RADICAL, dotted lines of RADMC, and dashed lines of STEINRAY from the
      semi-analytical solution.}
      \label{fig:diffopth}
   \end{figure}  
\subsection{Approximate solution for optically thin configurations}{\label{sect:semi-ana}}
        In the case of configurations being optically thin for all relevant
        wavelengths, heating of the dust particles is dominated by the stellar
        radiation. 
        When re-emission of the dust particles can be neglected and
        scattering is only included as extinction term, the 
        dust temperature can be easily determined from eq. (\ref{enerbal}) 
        without any coupling to eq. (\ref{radtrans}).
        We use this approximate semi-analytical solution to test independently 
        our RT computations for the most optically thin case and to check 
        the correctness of the density setup
        in the other more optically thick models (see Sect. \ref{sect:temp} and \ref{sect:sed}).
         
        The assumptions we discussed above simplify eq. (\ref{enerbal}) in the following way
\begin{equation}\label{eq:tem}  
        \int_{\lambda_{\mathrm{min}}}^{\lambda_{\mathrm{max}}}d\lambda\ 
        B_\lambda[T_{\mathrm{d}}(R,\theta)]  Q^{\mathrm{abs}}_\lambda   =
        \left(\frac{R_*}{2 R}\right)^2 
        \int_{\lambda_{\mathrm{min}}}^{\lambda_{\mathrm{max}}}d\lambda\ B_\lambda^e(T_*,R,\theta)  
         Q^{\mathrm{abs}}_\lambda    
\end{equation}
        with $T_{\mathrm{d}}(R,\theta)$ being the disk temperature at the location 
        $(R,\theta)$\footnote{$R$ is the distance from the central star in spherical coordinates and $\theta$
        is the polar angle as measured from the disk midplane} and $Q^{\mathrm{abs}}(\lambda)$ 
        the absorption efficiency factor. We perform the integration at the same wavelengths
        as those adopted by the RT codes (see Sect. \ref{defmodel}).
        The term $B_\lambda^e(T_*,R,\theta)$ represents the black body emission 
        from the star corrected for the extinction
\begin{equation}\label{eq:tau}  
        B_\lambda^e(T_*,R,\theta) = B_\lambda(T_*) \, 
        e^{-\pi a^2 (Q^{\mathrm{abs}}_\lambda+Q^{\mathrm{sca}}_\lambda)\int_{\mathrm{0}}^{\mathrm{R}}dR'\rho(R',\theta)}
\end{equation}  
        where $a$ is the dust radius, $Q^{\mathrm{sca}}(\lambda)$ is the scattering efficiency factor and
        $\rho(R',\theta)$ is the density distribution given in eq. (\ref{eq:den}) but here expressed in 
        spherical coordinates. The argument of the exponent represents the optical depth $\tau_\lambda(R,\theta)$
        at the distance $R$, $\theta$ from the central star. 
        Once the optical depth is determined, the extincted black body emission can be substituted in eq.(\ref{eq:tem})
        and the dust temperature can be easily computed.        
        To obtain the flux density $F_{\lambda}$ at a distance equal to
        the stellar radius we need to integrate the power emitted by each grain over the entire volume.
        If we express the power emitted by one grain as
\begin{equation}\label{eq:gemis}
        P_\lambda^{\mathrm{g}}(R,\theta) = 4 \pi a^2 \, Q^{\mathrm{abs}}_\lambda  B_\lambda[T_{\mathrm{d}}(R,\theta)]
\end{equation}  
        the flux density can be obtained by solving the following integrals
\begin{equation}\label{eq:sed1}
         F_\lambda = \frac{2\pi}{4 \pi R_*^2} \int_{\mathrm{R_{in}}}^{\mathrm{R_{out}}} \int_{\mathrm{\theta=0}}^{\mathrm{\pi}} 
         d\theta'   dR'
         P_\lambda^{\mathrm{g}}(R',\theta')\, \rho(R',\theta')\, R'^2 cos(\theta') 
\end{equation}
        here the factor $2 \pi$ comes from the integration in $\phi$ which is the azimuthal angle in the $x-y$ plane.
        
        A first rough estimate of the correctness of this approach can be done by evaluating how much 
        the disk temperature $T_{\mathrm{d}}$ would increase because of secondary emission re-absorption events.
        The grain emissivity is proportional to $T_{\mathrm{d}}^4$ and, in case of optically thin emission,
        to the optical depth. 
        Our most optically thin model has a maximum IR optical depth of 
        0.01 at 10~\micron{} in the disk midplane, where most of the dust is confined.
        If an emitted IR photon were re-absorbed by the disk, its temperature would increase by
        the quantity $(1+\tau_{\mathrm{IR}})^{0.25}$. 
        Thus, the temperature obtained by the semi-analytical approach can be considered correct within $0.26$\%
        in a first approximation.
        The corresponding emergent flux has an uncertainty which is about four times larger (about $1$\%).
        However looking at the SED differences (see Fig. \ref{fig:diffopth}), it is clear that deviations due to 
        scattering (treated correctly in the numerical codes) are important too.
        A more realistic estimate of the error on the semi-analytical approach should indeed
        consider the effect of scattering apart from damping the stellar flux. 
        
   \begin{figure}
   \resizebox{\hsize}{!}{\includegraphics[angle=0,width=2.5cm,height=2.5cm]{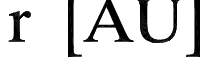}}
      \caption{Radial temperature (upper panel) and percentage of difference (lower panel) among the codes
      for the most optically thick model. 
      RADICAL is taken as reference code. The radial cut is made for an angle $\theta$ near to the disk midplane.
      Diamonds give the radial dependence in case of long wavelengths and optically thin emission.
      In the upper panel, solid lines are the results from MC3D, dot-dashed lines from MCTRANSF, 
      dashed-dot-dot-dot from RADICAL, dotted lines from RADMC and dashed lines from STEINRAY.
      In the lower panel, solid lines give the difference of MC3D, 
      dot-dashed lines of MCTRANSF, dotted lines of RADMC and dashed lines of STEINRAY from RADICAL.}
      \label{fig:tem100r}
   \end{figure}
   \begin{figure}
   \resizebox{\hsize}{!}{\includegraphics[angle=0,width=2.5cm,height=2.5cm]{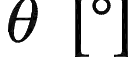} }
      \caption{Vertical temperature and percentage of difference among the codes
      for the most optically thick model and for a distance $r$ in the midplane equal to 2~AU from the central star. 
      RADICAL is taken as reference code. The nomenclature is the same as for Fig. \ref{fig:tem100r}
      except for the diamonds which give the temperature behaviour following the semi-analytical approach 
      described in Sect. \ref{sect:semi-ana}. Note that all the RT codes have the same turnover in the 
      temperature distribution at the location predicted by the semi-analytical solution.} 
      \label{fig:tem100v}
   \end{figure}                 
\subsection{Temperature}{\label{sect:temp}}        
        All the codes correctly reproduce the shape of the temperature distribution, both
        its radial and vertical dependence.
        
        In oder to test independently the most optically thin case (highest optical depth
        in the disk midplane of $\tau_{\mathrm{v}}$~=~0.1), 
        we use the semi-analytical solution derived in Sect.~\ref{sect:semi-ana}.
        The upper panel of Fig.~\ref{fig:diffopth} shows the percentage of 
        difference\footnote{with difference we mean: $\frac{(a-b)}{b}$. 
        Here $b$ stands for the reference solution/code while $a$ for any other code.} 
        between the semi-analytical solution and any other code.         
        Radial cuts are plotted for an angle $\theta$ near to the disk midplane where
        the influence of scattering is the largest.
        The differences between the semi-analytical solution and the RT codes is always smaller than
        1\%. 
          
        Temperature distributions for the models with $\tau_{\mathrm{v}}$~=1 and 10 agree better than
        10\% for all the cases we examine.
        The disk model with the highest optical depth is the most difficult to treat for the RT codes.
        In Fig. \ref{fig:tem100r} and  \ref{fig:tem100v}, we show radial and vertical cuts
        at the disk locations where deviations from the codes are expected to be higher, i.e.
        near to the disk midplane and close to the central star. 
        The radial temperature is plotted for an angle $\theta$ equal to 2.5$^{\circ}$
         from the disk midplane while the vertical temperature is given for a distance equal to 2~AU
         from the central star.  
        In the upper panel of Fig. \ref{fig:tem100r}, we also superimpose the
        temperature dependence for the optically thin regime at long wavelengths.
        In this regime the temperature distribution  depends only on the dust properties and 
        can be approximated by
$T(r)\propto r^{-2/(4+\beta)}$ (Evans \cite{evans94}).
        Here $\beta$ corresponds to the index of the dust absorption 
        coefficient at low frequencies ($\kappa_{\nu}^{\mathrm abs}\propto \nu^{\mathrm \beta}$).
        For Draine \& Lee silicates $\beta$ is equal to 2, leading to an 
        exponent of $-0.33$ in the temperature relation. 

        The upper panel of Fig.~\ref{fig:tem100v} provides (in diamonds) the vertical temperature
        profile from the semi-analytical approach described in Sect.~\ref{sect:semi-ana}.
        The semi-analytical solution has the turnover point from optically thick to
        optically thin (the place where the temperature suddenly starts to drop) 
        around $19\degr$ from the midplane. Since the solution provided by the RT codes should
        have the same behaviour, we used the semi-analytical approximation to check the correctness
        of the density setup. 
        The disk midplane calculated with the semi-analytical approach is naturally cooler 
        than the real disk because the approximation neglects heating of the disk from dust re-emission.
        The outer regions of the real disks are also warmer because of those photons scattered far from
        the midplane.
        At a distance $r$ of 2~AU from the star and exactly in the midplane the real temperature is 
        a factor of about 1.3 higher than that given by the semi-analytical solution. 
\begin{figure*}
        \resizebox{\textwidth}{!}{\includegraphics[angle=0]{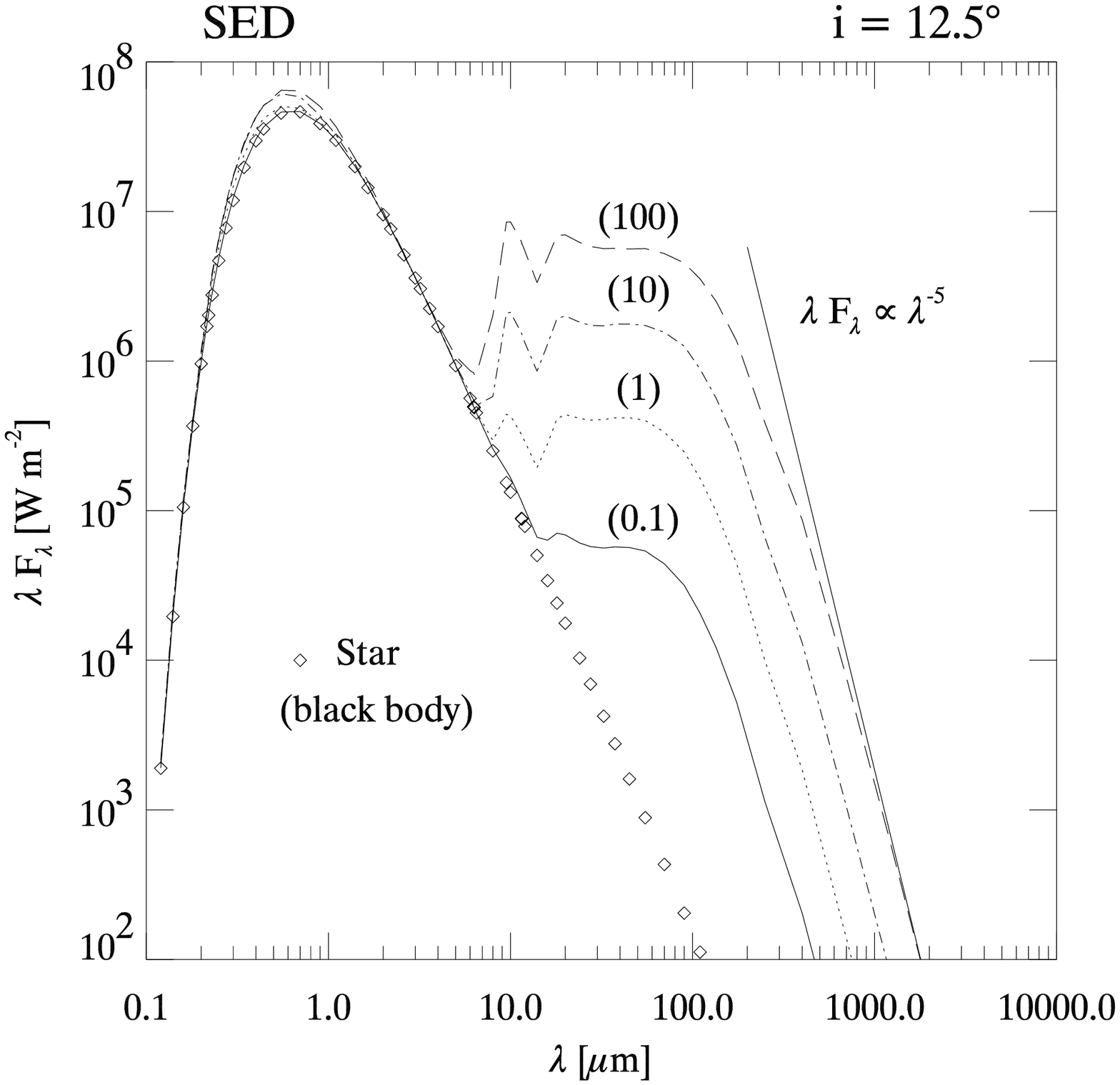}%
                                  \includegraphics[angle=0]{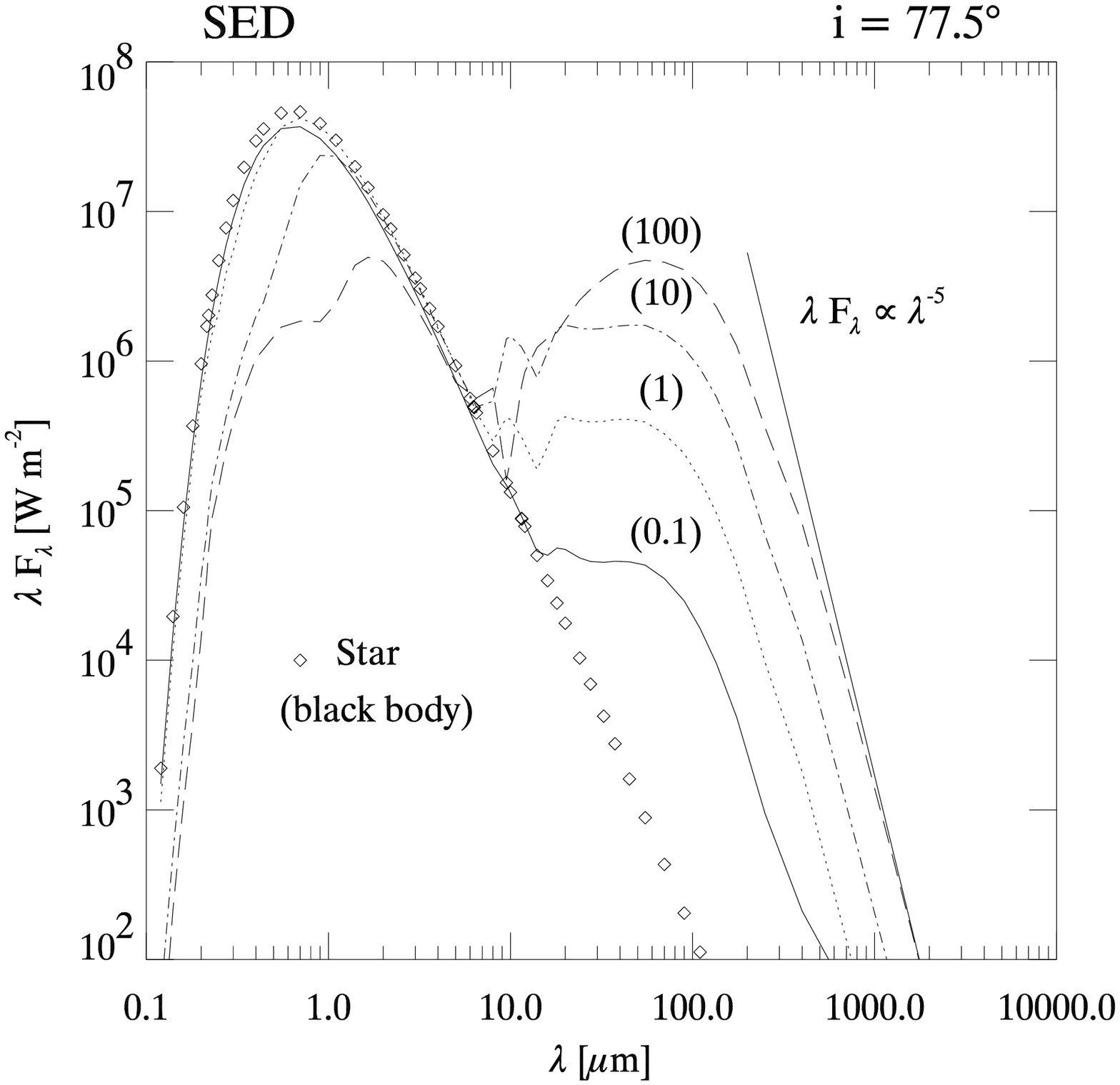} }
        \caption[]{SED for two disk inclinations $i$ as given on top of each panel.
        Each curve provides the mean value of the five RT simulations for the
        four computed models with different optical depth. The midplane optical
        depth is given in parenthesis labeling each curve.
        In both panels solid lines show results for the most optically thin disk,
        dotted lines for a disk having $\tau_{\mathrm{v}}=1$, dot-dashed lines for a
        disk with $\tau_{\mathrm{v}}=10$ and dashed lines for the most optically thick model.
        Diamonds provide the black-body emission from the naked star.
        The slope of the SED at long wavelengths depends only on the dust properties and is plotted
        in each panel with a solid line.} 
        \label{fig:sed}
\end{figure*}
        
        The lower panels of Fig.~\ref{fig:tem100r} and \ref{fig:tem100v} provide the percentage of difference among the codes
        taking RADICAL as reference code.
        The radial temperatures agree better than 5\% in most of the disk, going from 1.2 AU to
        200 AU.  Around 1.1 AU MCTRANSF deviates slightly more than 10\%. 
        STEINRAY shows 10\% deviations at the
        inner boundary and slightly higher deviations (but always less than 15\%) far from the star, 
        at about 900 AU. The vertical cut at 2 AU shows an agreement better than 2.5\% till 10$\degr$ from the
        disk midplane. Closer to the disk midplane, deviations are larger for MC3D and STEINRAY but always smaller than
        4\%.   

\subsection{Spectral Energy Distribution}{\label{sect:sed}}
        The emerging spectra for the four models having different optical depths
        are shown in Fig. \ref{fig:sed} at two disk inclinations.
        The left panel provides the results for an almost face-on disk 
        (disk inclination $i$ equal to 12.5$\degr$) while the right
        panel gives the result for an almost edge-on disk ($i=77.5\degr$).      
        Each curve represents the mean value of the five RT simulations for the specific
        model, whose optical depth is given in parenthesis above the curve. 
        On the y axis, we plot  $\lambda\,F_{\lambda}$ in [W\,m$^{-2}$]
        where $F_{\lambda}$ is the flux density at a distance equal to
        the star radius.        
        We also superimpose in diamonds the black body radiation arising 
        from the star in order to visualize how efficiently the circumstellar 
        disk reprocesses the stellar energy. 
        We note that all the codes have the correct slope at long wavelengths.
        This slope depends only on the dust properties and is plotted as solid line
        in both panels ($\lambda F_\lambda \propto \lambda^{-5}$). 
        At 0.55~\micron{} the drop in luminosity amounts to about a factor of 20
        going from the most optically thin to the most optically thick model 
        and for a disk inclination of 77.5$^{\circ}$.
\begin{figure*}
        \resizebox{\textwidth}{!}{\includegraphics[angle=90,width=2.5cm,height=2.5cm]{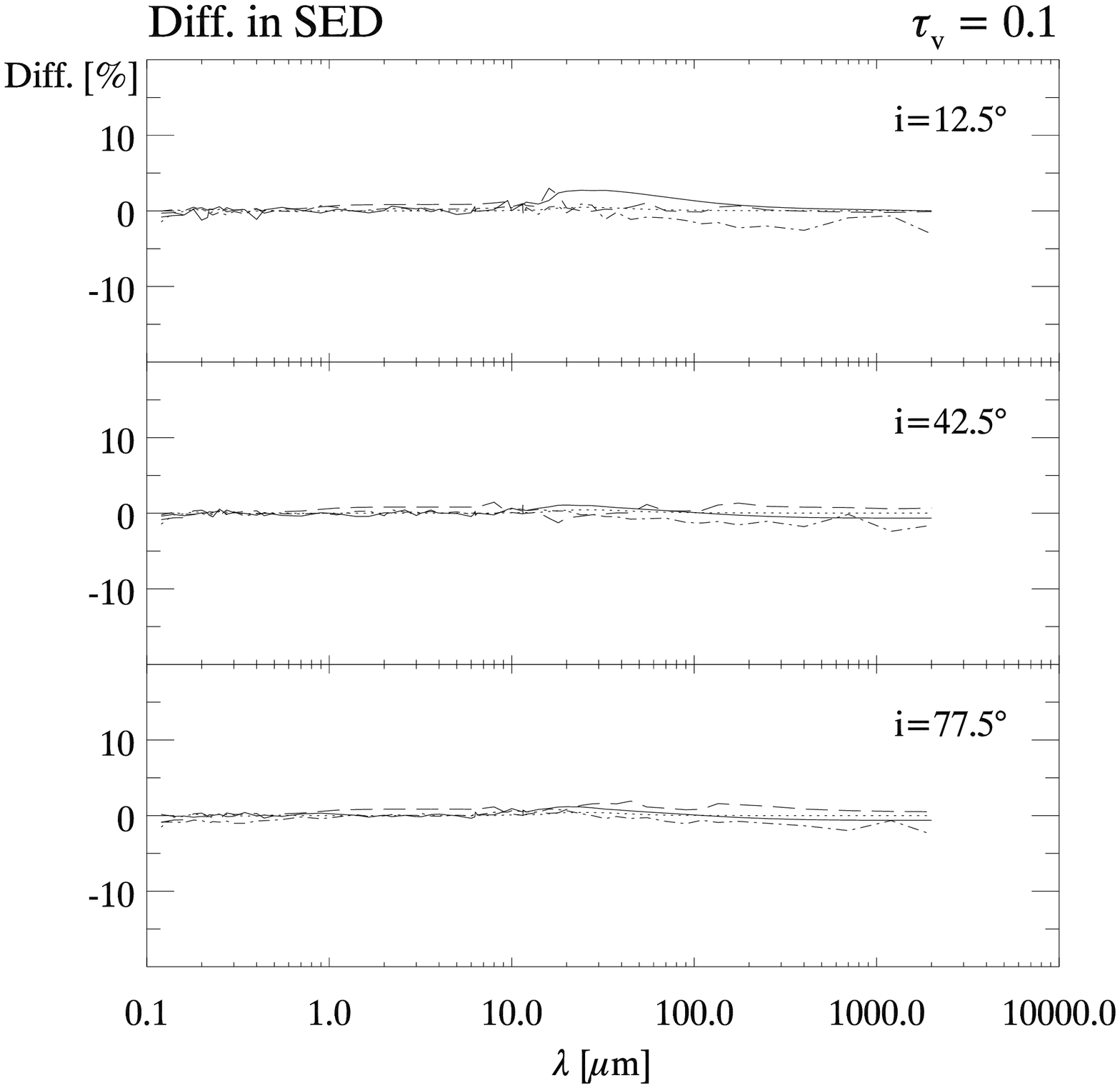}%
                              \includegraphics[angle=90,width=2.5cm,height=2.5cm]{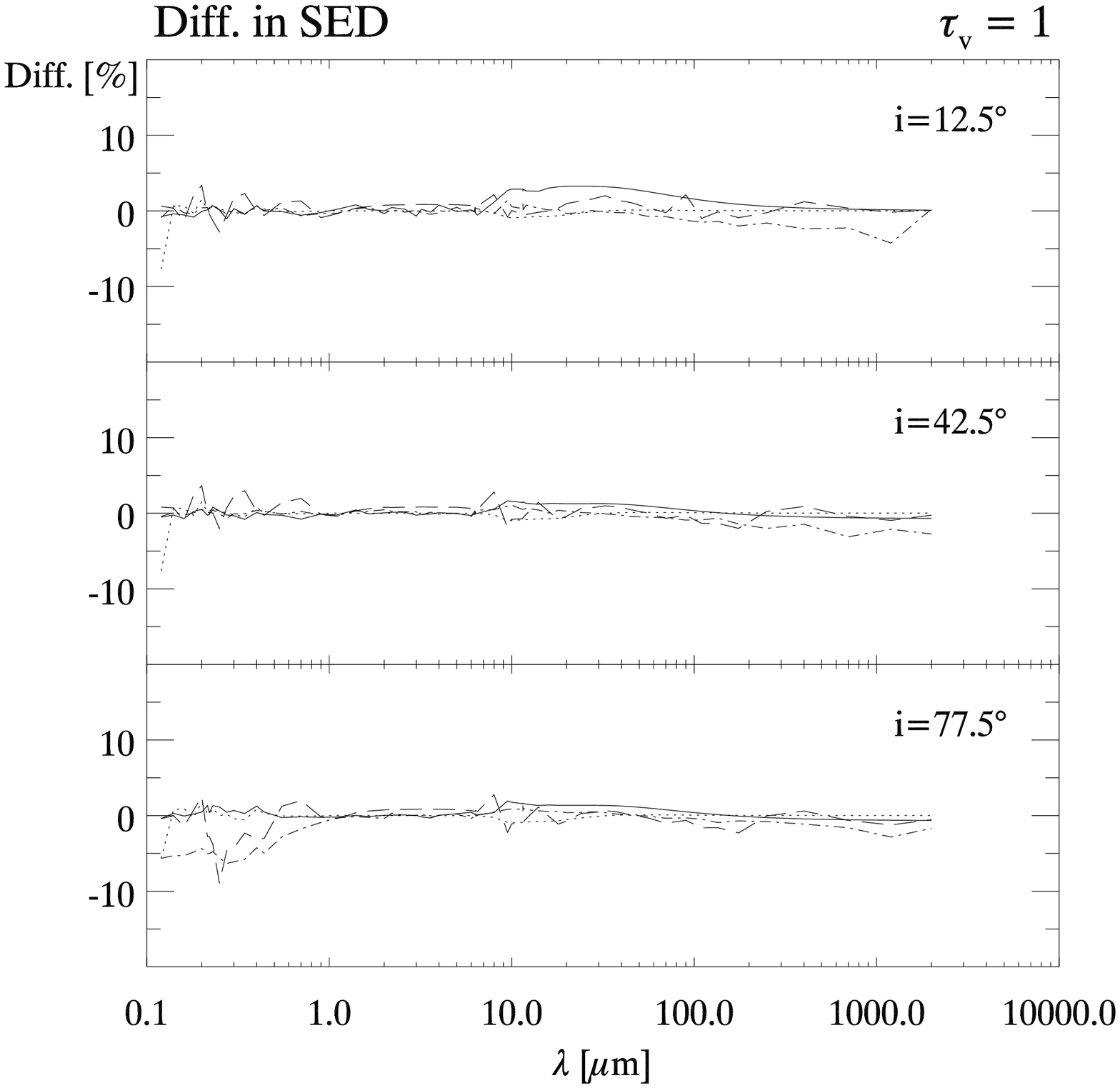}}
        \resizebox{\textwidth}{!}{\includegraphics[angle=90,width=2.5cm,height=2.5cm]{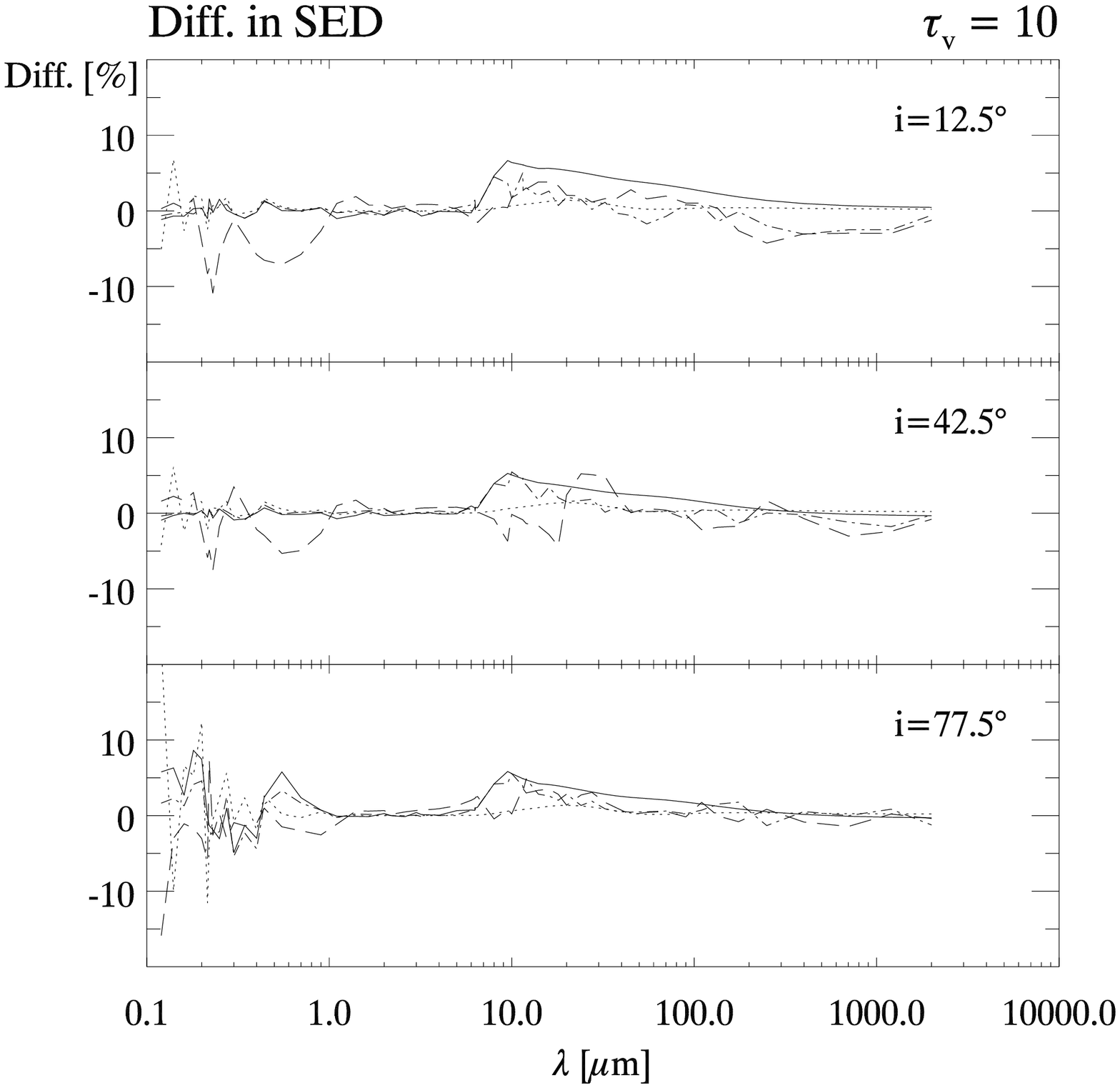}%
                              \includegraphics[angle=90,width=2.5cm,height=2.5cm]{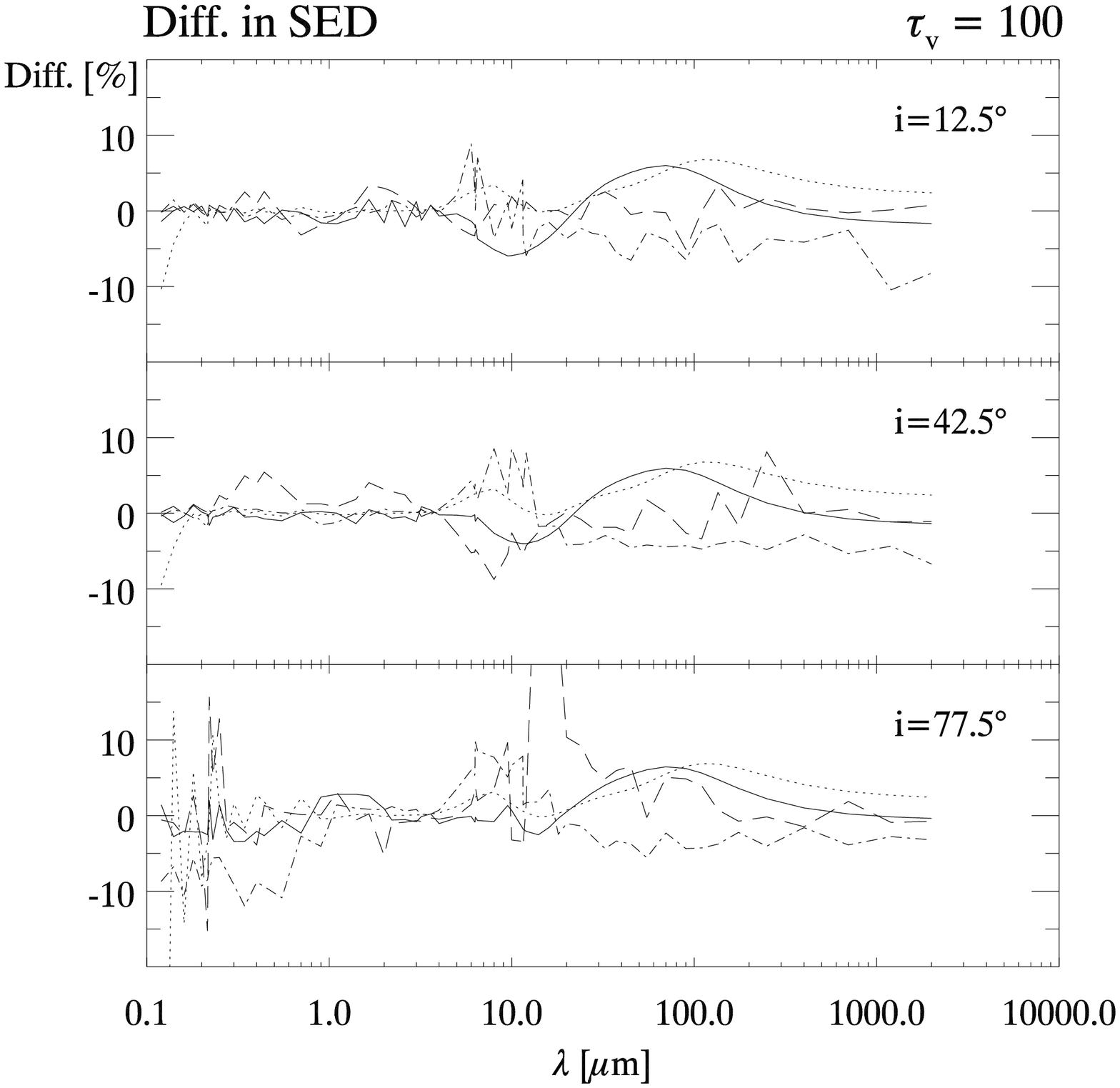}}
        \caption[]{Percentage of difference in the SED between the codes. RADICAL is taken as reference code. 
        Solid lines give the difference between MC3D and RADICAL, 
        dot-dashed lines between MCTRANSF and RADICAL, dotted lines
        between RADMC and RADICAL and dashed lines between STEINRAY and RADICAL.} 
        \label{fig:diffsed}
\end{figure*}  
        
        Since the differences among the codes are too small to be visible in a logarithmic
        plot, we provide separately  the percentage
        of difference for the four models and for three disk inclinations
	(see Fig. \ref{fig:diffsed}).
        RADICAL has been chosen as reference code. 
        For the most optically thin case, we also compare our results with the semi-analytical approach
        (see Fig. \ref{fig:diffopth}).
	We find that the agreement of the codes with the semi-analytical solution is always 
        better than 8\%, with the largest deviations around  0.3 and 40~\micron . 
        In the range 0.2--0.7~\micron{} all the codes predict higher flux in comparison to the
        semi-analytical solution, while between 10--200~\micron{} a  lower flux is obtained.
        These deviations arise because the semi-analytical approach includes 
        scattering only as an extinction term. 
        From the numerical RT calculations it is clear that some photons are scattered
        thus enhancing the flux  between 0.2 and 0.7~\micron . 
        We note that this wavelength range is exactly where small astronomical
        silicate grains have the largest scattering efficiency (see Fig.~\ref{fig:opt}).
        Therefore, deviations peaking at 0.3~\micron{} are simply explained 
        by the particular optical data chosen for this benchmark.
        Those photons which are scattered cannot contribute
        to heat the disk. This explains why RT codes predict a lack of emission 
        at longer wavelengths. To understand why the largest deficit of photons is around 40--50~\micron{},
        we first compute the temperature at which most of the disk mass emits 
        (mass average temperature) and then the corresponding wavelength.
        For the wavelength calculation we need to take into account the grain emissivity
        (Evans \cite{evans94}). We find a mass average temperature of 40~K which translates into a wavelength
        of 50~\micron{} at the maximum emission. This wavelength is well in agreement with the
        deviations shown in Fig.~\ref{fig:diffopth}.  
        For comparison, the RT codes agree better than 1.5\% at wavelengths shorter
	than 10~\micron{} for this particular test case ($\theta = 12.5\degr$). 
        At longer wavelengths the results show a bit more scatter 
        but the agreement is always better than 3\%.
        In Fig.~\ref{fig:diffsed} first panel, we also show the percentage of difference for
        two other disk inclinations, namely 42.5$\degr$ and 77.5$\degr$.
        In both cases the agreement is better than 2\% at all wavelengths for all the codes
        but MCTRANSF, for which slightly higher deviations  (about 2.5\%) 
        are present at longer wavelengths.  
        
        As the optical depth in the midplane increases, the RT problem becomes more difficult to solve.
        Because of the chosen disk geometry, most of the disk
        mass is located near to the midplane and close to the disk inner boundary.
        Our comparison shows that agreement among the codes is always better
        for an almost face-on case and 42.5$\degr$ disk inclination
        (first two panels of Fig.~\ref{fig:diffsed}), than for an almost edge-on
        disk (lower panels of Fig.~\ref{fig:diffsed}).
        For the models with $\tau_{\mathrm{v}}=$1 deviations among the codes are smaller than 9\%.
        For the disk with midplane optical depth of 10 and 100
        and inclinations of 12.5 and 42.5$\degr$, differences do not exceed 10\%.
        For the almost edge-on configurations the most difficult  regions to treat are
        those where scattering dominates and at wavelengths around 10~\micron . 
        In the IR, opacities change strongly and the modified Planck emission 
        peaks in the inner disk regions (between 1 and 2~AU).
        Therefore, the numerical simulations are particularly sensible to the resolution of 
        the inner parts. 
        Deviations in the IR are partly due to the different resolution adopted by the codes
	(see also Sect.~\ref{sect:tests}).
        Scatter at visible and near-infrared wavelengths for MC3D, MCTRANSF, and RADMC is simply
        statistical noise, typical of MC simulations. 
        This scatter becomes more prominent at high optical depths.
	We also note that MC3D and RADMC have the same trend for wavelengths larger than 10~\micron{}
	and the model with $\tau_{\mathrm{v}}=$100:
	they both estimate a larger IR emission than RADICAL with peaks at $\sim$70~\micron{} for
	MC3D and at $\sim$100~\micron{} for RADMC. 
        A strong deviation from the other codes is shown by STEINRAY just after the 9.8~\micron{} silicate 
        feature. 
        However, one should note that apart from the discussed features the overall agreement of the SEDs
        is better than 10\% for all the codes even for the almost edge-on disk and the most optically thick 
	test case.        
\subsection{Tests for various spatial and frequency resolutions}
\begin{figure}
        \resizebox{\hsize}{!}{\includegraphics[angle=90]{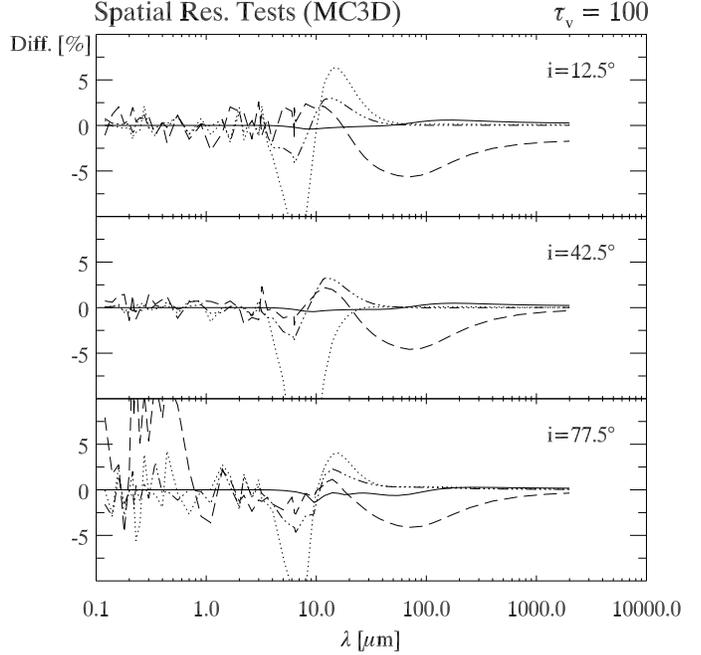}}
        \caption[]{
        Spatial resolution tests using the MC code MC3D. On the y-axis we plot
        the percentage of difference between the emerging SED of {\it mod0} and any other model
        in Table~\ref{tab:testsp}. Solid line: difference between {\it mod0} and {\it mod1}.
        Dot-dot dahsed line: difference between {\it mod0} and {\it mod2}. 
	Dotted line: difference between {\it mod0} and {\it mod3}.
	Dashed line: difference between {\it mod0} and {\it mod4}.
        } 
        \label{fig:testsp}
\end{figure}

{\label{sect:tests}}
       We used the MC code MC3D to test the dependence of our results on the grid
       adopted to store the emerging temperature. 
       Since deviations due to different temperature sampling are expected to be larger
       for more optically thick configurations, we investigate our most optically thick test case
       $\tau_{\mathrm{v}}$~= 100. Different grids, covering radial resolutions
       from 2.7~AU up to 0.03~AU in the inner disk and vertical resolutions from 1.5$\degr$ to 5.8$\degr$,
       have been inspected.
       In Fig.~\ref{fig:testsp}, we report results for five relevant cases. The 
       number of radial and vertical subdivisions (\#$r$ and \#$\theta$), as well as the resolution ($\Delta$r and
       $\Delta \theta$) and total number of photons (\#Phot) for these cases
        are provided in Table~\ref{tab:testsp}. 
       The RT equation is solved for 61 wavelengths, the same assumed in all the previous simulations.
       The number of photons is set to 4$\times$10$^5$ per wavelengths to limit the runtime to 2 days
       on a PC with 4~Gby memory, 2.4~GHz clock. Only in one case ({\it mod4}) we let the
       computation run for about two weeks in order to reduce as much as possible the photon noise at short
       wavelengths. The comparison  shows that the vertical spacing does not influence too much the
       results: we report deviations smaller than 1.5\% for the three disk inclinations between the models 
       with vertical resolution 1.5$\degr$ and 5.8$\degr$ (solid line in Fig.~\ref{fig:testsp}). 
       On the other side, changes in the radial grid strongly affects the IR emission:
       when going from {\it mod0} to {\it mod2} results still do not differ more than 5\% but the coarse inner 
       grid of {\it mod3} causes deviations larger than 20\%.
       Model {\it mod4} is the state-of-the-art for our computer capabilities.
       The grid has a good resolution also in the outer region of the disk and we use 10 
       times more photons per wavelengths.
       The percentage of difference between this model and {\it mod0} in the IR regime  
       amounts to less than 6\%. 
       Deviations of about 10\% in the optical range  are due to differences in the number of photons.
       
   \begin{table}
      \caption[]{Relevant models for the spatial resolution tests}
         \label{tab:testsp}
     $$ 
         \begin{array}{p{0.1\linewidth}p{0.1\linewidth}p{0.2\linewidth}p{0.1\linewidth}p{0.1\linewidth}p{0.1\linewidth}}
            \hline
            \hline
            \noalign{\smallskip}
            Model       & \# r   & $\Delta$r& \# $\theta$&$\Delta \theta$ & \#Phot\\
                        &       & [AU]     &          & [ $\degr$ ]     & [$\times$10$^6$] \\   
            \hline
            {\it mod0}  & 55   & 0.03--141&   101     & 1.8  & 24.4\\
            {\it mod1}  & 55   & 0.03--141&    31     & 5.8  & 24.4\\
	    {\it mod2}  & 40   & 0.3--141 &   121     & 1.5  & 24.4\\
            {\it mod3}  & 35   & 0.7--141 &   121     & 1.5  & 24.4\\
            {\it mod4}  & 10$^3$ & 0.07--4.1&   121     & 1.5  & 244\\
            \noalign{\smallskip}
            \hline
         \end{array}
     $$ 
   \end{table}         
       
\begin{figure}
        \resizebox{\hsize}{!}{\includegraphics[angle=90]{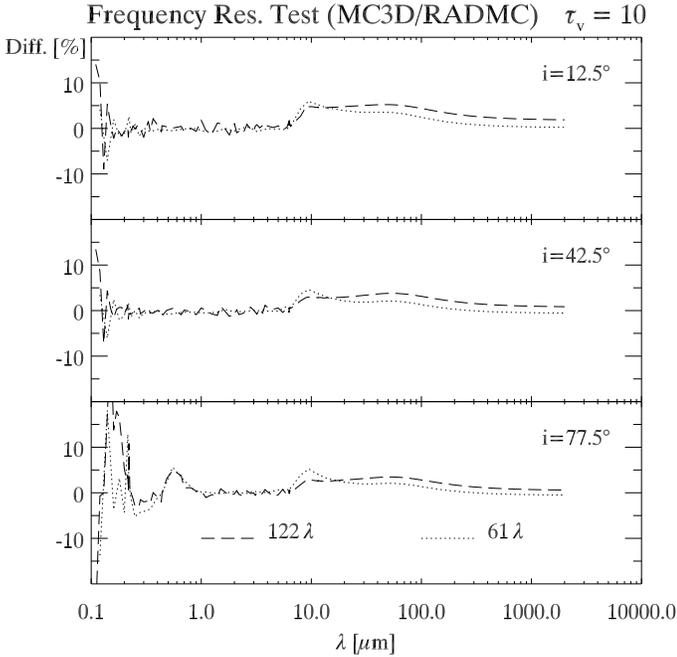}}
        \caption[]{
        Frequency resolution test using the MC codes MC3D and RADMC. On the y-axis we plot
        the percentage of difference between the emerging SED of the codes for two models,
        one sampling the frequency space with 61 (dotted line) and the other with 122 (dashed line)
        logarithmically distributed points.
        } 
        \label{fig:testfreq}
\end{figure}

        To test the influence of the frequency resolution on our results we run the
        2 MC codes MC3D and RADMC doubling the number of walenghts where to solve the RT equation. 
        The other three codes could not take part to the comparison because of not enough 
        computer memory.
	To have reasonable runtime for MC3D (a couple of days), we restrict ourselves to
	the case $\tau_{\mathrm{v}}$~= 10 and we lower the number of photons
	in comparison to Table~\ref{tab:simu} (half in the case of RADMC and 4 times lower in the case of MC3D).
	The lower number of photons introduces larger scatter 
        at short wavelengths. Fig.~\ref{fig:testfreq} shows the percentage of difference
        between MC3D and RADMC for the model with 61 wavelenghts (dotted line) and the new model with
        122 wavelenghts (dashed line). Apart from the expected larger deviations at short wavelengths,
        the agreement between the two codes improves not more than 2\% around 10\micron .
        Thus, we conclude that the nature of the IR deviations plotted in 
        Fig.~\ref{fig:diffsed}, is not due to coarse frequency sampling but
	to the different grid resolutions adopted by the codes 
	(especially in radial direction) together with cumulative numerical errors.

\section{Discussion and Conclusions}

        Before presenting our findings, we briefly discuss the general features 
        of the computed SEDs for different viewing angles ($i$) and optical 
        depths ($\tau$). As mentioned in Sect. \ref{defmodel}, optical depths
        are measured in the disk midplane from the inner to the outer boundary. 
        Thus, the optical depths we refer to are the highest in the disk.

        Both panels of Fig. \ref{fig:sed} show clearly that
        the far-infrared region (longward 100~\micron ) remains unaffected 
        when varying the viewing angle. On the other hand, the short-wavelength part of
        the spectrum is strongly modified.
        When the disk is seen face-on the spectrum is dominated
        by unattenuated stellar radiation.
        As the disk inclination increases, more and more of the stellar flux is extincted 
        by the dust in the disk. For the $\tau_{\mathrm v}=100$ case at $i=77.5\degr$ this reduction amounts 
        to a factor of  $e^{-100}$ in the visual. However, due to the
        high albedo of the dust grains, a large fraction of the stellar radiation is scattered 
        above the disk into the line of sight. Dust scattering is also responsible for the
        excess of emission at stellar wavelengths seen at small disk inclinations
        (see left panel of Fig. \ref{fig:sed} especially for high optical depths).      
        The optical depth also affects the strong 10~\micron{} feature produced by the
        SI -- O stretching. While in most cases the feature appears strongly in emission, 
        for the $\tau_{\mathrm v}=100$ test case at $i=77.5\degr$ 
        the feature appears in absorption (see Fig. \ref{fig:sed} right panel).         
        The 20~\micron{} feature is much weaker than the 10~\micron , but it is already
        visible for the model with $\tau_{\mathrm v}$ = 1. 
        All these features are in agreement with earlier radiative transfer computations of disks 
        (e.g.~Efstathiou \& Rowan-Robinson \cite{efstathiou90},  Menshchikov \& Henning \cite{mensh97}). 

        Our aim in this paper is to provide benchmark solutions for the 2-D continuum
        radiative transfer problem in circumstellar disks. The problems we present have 
        optical depths up to 100, which is actually the limit of current computational capabilities
	for most of the codes. 
        The corresponding total dust mass in the disk of about 0.01 solar masses
        covers most of the observed disks around low mass stars.
        For more massive disks around intermediate and high mass stars as well
        as tori obscuring active galactic nuclei, the numerical strategies have
        to be modified, using e.g. the diffusion approximation for high optical
        depths.
        We used five independent radiative transfer codes that implement  different
        numerical schemes. 
        We compare both the resulting temperature structure and the emerging SEDs. 
        For the lowest optical depth case ($\tau_{\mathrm v}=0.1$) we also compared the results 
        against a semi-analytic solution which treat scattering only as extinction term.
        The other three cases ($\tau_{\mathrm v}=1, 10, 100$) cannot be solved in a
        semi-analytic way, since multiple scattering and absorption-reemission events
        strongly affect the solution. 
        
        We find that the overall shape of the temperature distribution and of the emerging
        SEDs is well reproduced by all the codes.
        Differences in the temperature are smaller than 1~\%  for all the codes in the most 
        optically thin case.
        Even for the most optically thick model, differences in the temperature remain below
        $15$\%.
        As for the SEDs, deviations among the codes are smaller than 3\% at all
        wavelengths and disk inclinations for the most optically thin model.
        For the models with $\tau_{\mathrm v}=1$ and 10 at all disk inclinations
        and for the most optically thick case for disk inclinations of 12.5 and 42.5$\degr$,
        differences do not exceed 10\%.
        Only for the most optically thick case and an almost edge-on disk
        differences around 10~\micron{} exceed 20\% in the case of 
        STEINRAY. 
	We stress that this is the case for which the numerics is
        the most difficult: the codes have to treat both a very optically thin atmosphere and
        a thick disk midplane. 
        Independent tests using two of the MC codes show that the frequency resolution
	cannot account for the infrared deviations among the codes 
        in the almost edge-on disk and the most optically thick model.
	Grid resolution especially in radial direction together with cumulative numerical errors 
	play a major rule. The presented results provide a robust way to test other
	continuum RT codes and demonstrate the possibilities of the current computational capabilities. 	 
        Temperature distributions and SEDs for all the test cases are available at the web site: \\
        http://www.mpia.de/PSF/PSFpages/RT/benchmark.html 
\begin{acknowledgements}
        The developements of MCTRANSF has taken advantages
        of several discussions with Anne Dutrey to adapt its performances in
        agreement with the problems related to the disk of YSOs.
        C. P. Dullemond acknowledges support from the European Commission under TMR grant
        ERBFMRX-CT98-0195 (`Accretion onto black holes, compact objects and
        prototars')

\end{acknowledgements}


\begin{thebibliography}{}

\bibitem[1996]{bachiller96} Bachiller, R. 1996, \araa, 34, 111

\bibitem[2001]{bjork01} Bjorkman, J. E., Wood, K. 2001, \apj, 554, 615

\bibitem[1959]{cashwell59} Cashwell, E.D., Everett, C.J. 1959,
        "A practical manual on the Monte Carlo Method for random walk problems."
        Pergamon, New York
        
\bibitem[1997]{cg97} Chiang, E. I. \& Goldreich, P. 1997, \apj, 490, 368

\bibitem[1999]{cg99} Chiang, E. I. \& Goldreich, P. 1999, \apj, 519, 279

\bibitem[1986]{chini86} Chini, R., Kr\"{u}gel, E., Kreysa, E. 1986, \aap, 167, 315

\bibitem[1991]{collfix91} Collison, A.J. \&  Fix, J.D. 1991, \apj, 368, 545

\bibitem[2001]{cotera01} Cotera, A. S., Whitney, B. A., Young, E. et al. 
                         2001, \apj, 556, 958

\bibitem[1984]{draine84} Draine, B. T. \& Lee, H. M. 1984, \apj, 285, 89 

\bibitem[2000]{dullemond00} Dullemond, C. P. \& Turolla, R. 2000, \aap, 360, 1187

\bibitem[2003]{dullemond03} Dullemond, C.~P. 2003, submitted to \aap

\bibitem[2002]{dulvzadnat02} Dullemond, C.~P., van Zadelhoff, G.~J., Natta, A. 2002, \aap, 389, 464

\bibitem[1990]{efstathiou90} Efstathiou, A. \& Rowan-Robinson 1990, \mnras, 245, 275

\bibitem[1991]{efstathiou91} Efstathiou, A. \& Rowan-Robinson 1991, \mnras, 252, 528

\bibitem[1994]{evans94} Evans, A. 1994, The Dusty Universe, 
                        Chichester, New York : J. Wile, p. 67-69 

\bibitem[1993]{groenewegen93} Groenewegen, M. A. T. 1993, \aap, 271, 180

\bibitem[1985]{henning85} Henning, Th. 1985, Astr. Space Sci., 114, 401

\bibitem[2001]{henning01} Henning, Th. 2001, The Formation of Binary Stars,
                                IAU Symposium, Vol. 200, p. 567


\bibitem[1971]{hummer71} Hummer, D. G., Rybicki, G. B. 1971, \mnras, 152, 1

\bibitem[1997]{ivezic97} Ivezic, Z., Groenewegen, M. A. T., Men'shchikov, A., Szczerba, R.
                         1997, \mnras, 291, 121
                         
                         
\bibitem[2002]{kikuchi02} Kikuchi, N., Nakamoto, T., Ogochi, K. 2002, PASJ, 54, 589

\bibitem[1988]{kunasz88} Kunasz, P. B. \& Auer, L. H. 1988, JQSRT, 39, 67

\bibitem[1982]{lefevre82} Lef\`evre, J., Bergeat, J., Daniel, J.Y. 1982, \aap, 114, 341

\bibitem[1987]{lenzen87} Lenzen, R. 1987, \aap, 173, 124

\bibitem[1976]{leung76} Leung, C. M. 1976, \apj, 209, 75

\bibitem[1995]{lopez95} Lopez, B., M\'ekarnia, D., Lef\`evre, J. 1995, \aap, 296, 752 

\bibitem[2000]{lopez00} Lopez, B. \& Perrin, J.-M. 2000, \aap, 354, 657

\bibitem[1999]{lucy99} Lucy, L.B. 1999, \aap, 344, 282

\bibitem[1991]{malbetbertout91} Malbet, F. \& Bertout, C. 1991, \apj, 383, 814
  

\bibitem[1984]{martin84} Martin, P. G., Rogers, C., Rybicki, G. B. 1984, \apj, 284, 317

\bibitem[1996]{mccau96} McCaughrean, M. J. \& O'Dell, C. R. 1996, \aj, 111, 1977

\bibitem[1997]{mensh97} Men'shchikov, A. B. \& Henning, T. 1997, \aap, 318, 879

\bibitem[1978]{mihalas78} Mihalas, D. M., Auer, L. H.,  Mihalas, B. R. 1978, \apj, 220, 1001

\bibitem[1984]{mihalmihal84} Mihalas, D. \& Mihalas, B. 1984, Foundations of radiation hydrodynamics (Oxford University Press)

\bibitem[2000]{natta00} Natta, A., Meyer, M. R., Beckwith, S. V. W. 2000, \apj, 534, 838



\bibitem[2003]{niccolini} Niccolini, G., Woitke, P., Lopez, B. 2003, \aap, 399, 703   

\bibitem[1984]{rogers84} Rogers, C., Martin, P. G. 1984, \apj, 284, 327

\bibitem[1980]{rowan80} Rowan-Robinson, M. 1980, \apjs, 44, 403

\bibitem[1991]{ryb91}  Rybicki, G.B. \&  Hummer, D.G. 1991, \aap, 245, 171

\bibitem[1976]{scoville76} Scoville, N. Z. \& Kwan, J. 1976, \apj, 206, 718


\bibitem[1996]{stein96} Steinacker, J., Thamm, E., Maier, U. 1996, JQSRT, 56, 1, 97

\bibitem[2002a]{stein02a} Steinacker, J., Hackert, R., Steinacker, A., Bacmann, A. 2002a, JQSRT, 73, 557

\bibitem[2002b]{stein02b} Steinacker, J., Michel, B., Bacmann, A., 2002b, JQSRT, 74, 183

\bibitem[2002c]{stein02c} Steinacker, J., Bacmann, A., Henning, Th. 2002c, JQSRT, 75, 765

\bibitem[2003]{stein03} Steinacker, J., Henning, Th., Bacmann, A., Semenov, D. 2003, \aap, 401, 405

\bibitem[1992]{stonemihnor92} Stone, J., Mihalas, D., Norman, M. 1992, \apjs, 80, 819

\bibitem[2002]{vanzadelhoff02} van Zadelhoff, G.-J., Dullemond, C. P., van der Tak et al. 2002, \aap, 395, 373  

\bibitem[1994]{winters94} Winters, J. M., Dominik, C., Sedlmayr, E. 1994, \aap, 288, 255

\bibitem[1998]{wolf98} Wolf, S., Fischer, O., Pfau, W. 1998, \aap, 340, 103

\bibitem[1999a]{wolf99a} Wolf, S., Henning, Th. 1999a, \aap, 341, 675

\bibitem[1999b]{wolf99b} Wolf, S., Henning, Th., Stecklum, B. 1999b, \aap, 349, 839

\bibitem[2000]{wolf00} Wolf, S. \& Henning, Th. 2000, Comp. Phys. Comm., 132, 166

\bibitem[2001]{wolf01a} Wolf, S., Stecklum, B., Henning, Th. 2001a, in ASP Conf. Ser. ,
                        IAU Symposium 200, 295, "The Formation of Binary Stars"
                        
\bibitem[2001]{wolf01b} Wolf, S. 2001b, \aap, 379, 690                  
                        
\bibitem[2002]{wolf02} Wolf, S., Voshchinnikov, N.V., Henning, Th. 2002, \aap, 385, 365 

\bibitem[2002]{wood02} Wood, K., Wolff, M. J., Bjorkman, J. E., Whitney, B. 2002, \apj, 564, 887

\bibitem[1981]{yorke81} Yorke, H. W., Shustov, B. M. 1981, \aap, 98, 125

\bibitem[1985]{yorke85} Yorke, H. W. 1985, Birth and infancy of stars, 
                (A86-31601 13-90). Amsterdam, North-Holland, p. 645
\end{thebibliography}
\end{document}